\documentclass{elsart}

\def\mathfont#1{\ifmmode{#1}\else{$#1$}\fi} 
\def\lae{\mathrel{<\kern-1.0em\lower0.9ex\hbox{$\sim$}}}  
\def\gae{\mathrel{>\kern-1.0em\lower0.9ex\hbox{$\sim$}}}  
\def\kms{\ifmmode{{\rm km\ s}^{-1}}\else{${\rm km\ s}^{-1}$}\fi} 

\def\ergsec{\mathfont{ {\rm ergs\ s}^{-1}}}

\def\msun{\ifmmode{\ {\rm M}_\odot}\else{$ {\rm M}_\odot$}\fi}  
\def\msunyr{\ifmmode{\msun \ {\rm yr}^{-1}}\else{$\msun \ {\rm yr}^{-1}$}\fi}

\def\und#1{$\underline{\smash{\hbox{#1}}}$}
 
\usepackage{harvard}

\usepackage{graphicx}

\usepackage{amssymb}

\def\url#1{{\ttfamily\def\/{/\discretionary{}{}{}}#1}}

\begin{document}
Presented at ``Life Cycles of Radio Galaxies'', July 15-17, 1999, STScI, Baltimore\\
\begin{frontmatter}
\title{Radio Triggered Star Formation in Cooling Flows}


\author[brm]{B.R. McNamara}, 

\thanks[brm]{E-mail: brm@cfa.harvard.edu}

\address[brm]{Harvard-Smithsonian Center for
Astrophysics, 60 Garden St. Cambridge, MA 02138}

\begin{abstract}
The giant galaxies located at the centers of cluster
cooling flows are frequently sites of vigorous star formation.
In some instances, star formation appears to have been triggered
by the galaxy's radio source.  The colors and spectral
indices of the young populations are generally consistent
with short duration bursts or continuous star formation
for durations $\ll 1$ Gyr, which is less than the presumed
ages of cooling flows.  The star formation properties
are inconsistent with fueling by a continuously
accreting cooling flow, although the prevalence of star formation
is consistent with repeated bursts and periodic refueling.
Star formation may be fueled, in some cases, by cold material
stripped from neighboring cluster galaxies.

\end{abstract}

\end{frontmatter}

\section{Introduction}

More than half of clusters within redshift $z\sim 0.1$ 
contain bright, central X-ray emission from $\sim$ keV gas that
appears to be cooling at rates of $\sim 10-1000 \msunyr$ (Fabian 1991).  
Commonly referred to as cooling flows, persistent accretion of 
this cooling material onto the 
bright, central galaxies in clusters (CDGs) at even a fraction of these
rates would be capable of fueling vigorous
star formation and the central engines generating their radio sources.
Enhanced levels of cold gas and star formation are indeed seen
in cooling flows (see McNamara 1997 for a review).  However,
the inferred star formation rates are only $\lae 1-10\%$ of
the cooling rates derived from X-ray observations, and the amounts 
of cold gas detected outside of the X-ray band would account for
$\lae 10^8$ yr of accumulated material.  
Between 60--70\% of CDGs in cooling flows harbor Type 1
Fanaroff-Reiley (FR I) radio sources, while
only $\sim 20\%$ of CDGs in non cooling flow clusters have bright
radio sources (Burns et al. 1997).  
The presence of a cooling flow increases dramatically
the likelihood of detecting a bright radio source in a CDG.
The radio sources in cooling flows are often interacting with
the cool (and hot) intracluster medium, 
influencing the gas dynamics (e.g. Burns et
al. 1997 and this conference), and in some cases, possibly triggering star
formation (McNamara 1997).
The origin of the cool material, whether
direct cooling from the intracluster medium, as would follow from
the standard cooling flow model, or an external source, such as
cold gas stripped from surrounding cluster galaxies, has not 
been identified conclusively.  I discuss these issues in this article,
illustrating several points with a brief analysis 
of new optical imagery for the Abell 1068 cluster CDG.

\section{Host Galaxy Properties}

A giant CDG resides at the base of all known
cooling flow clusters.  CDGs are, as a class, the largest and most 
luminous elliptical galaxies.
Their envelopes have been traced to radii of several hundred kiloparsecs
(Uson et al. 1991; Johnstone et al. 1991).  The absolute 
magnitudes of CDGs are typically $M_{\rm V}\sim -20$ to $-22$ within a 16 kpc radius
(Schombert 1987),  but they can be as luminous
as $M_{\rm I}\sim -26$ including the envelope (Johnstone
et al. 1991).

\begin{figure}
\begin{center}
\includegraphics*[width=7.5cm,height=6cm]{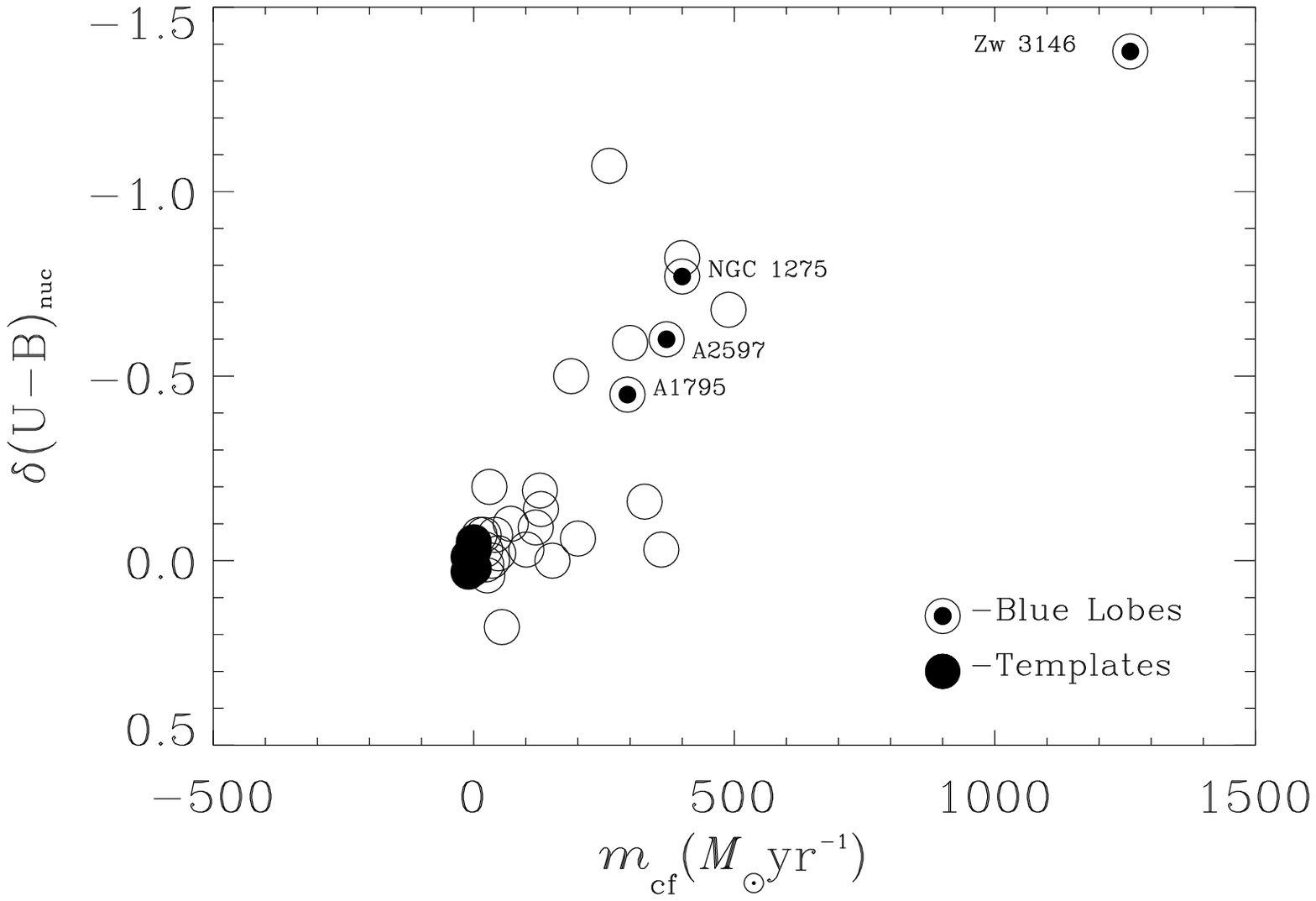}
\includegraphics*[width=6cm,height=6cm]{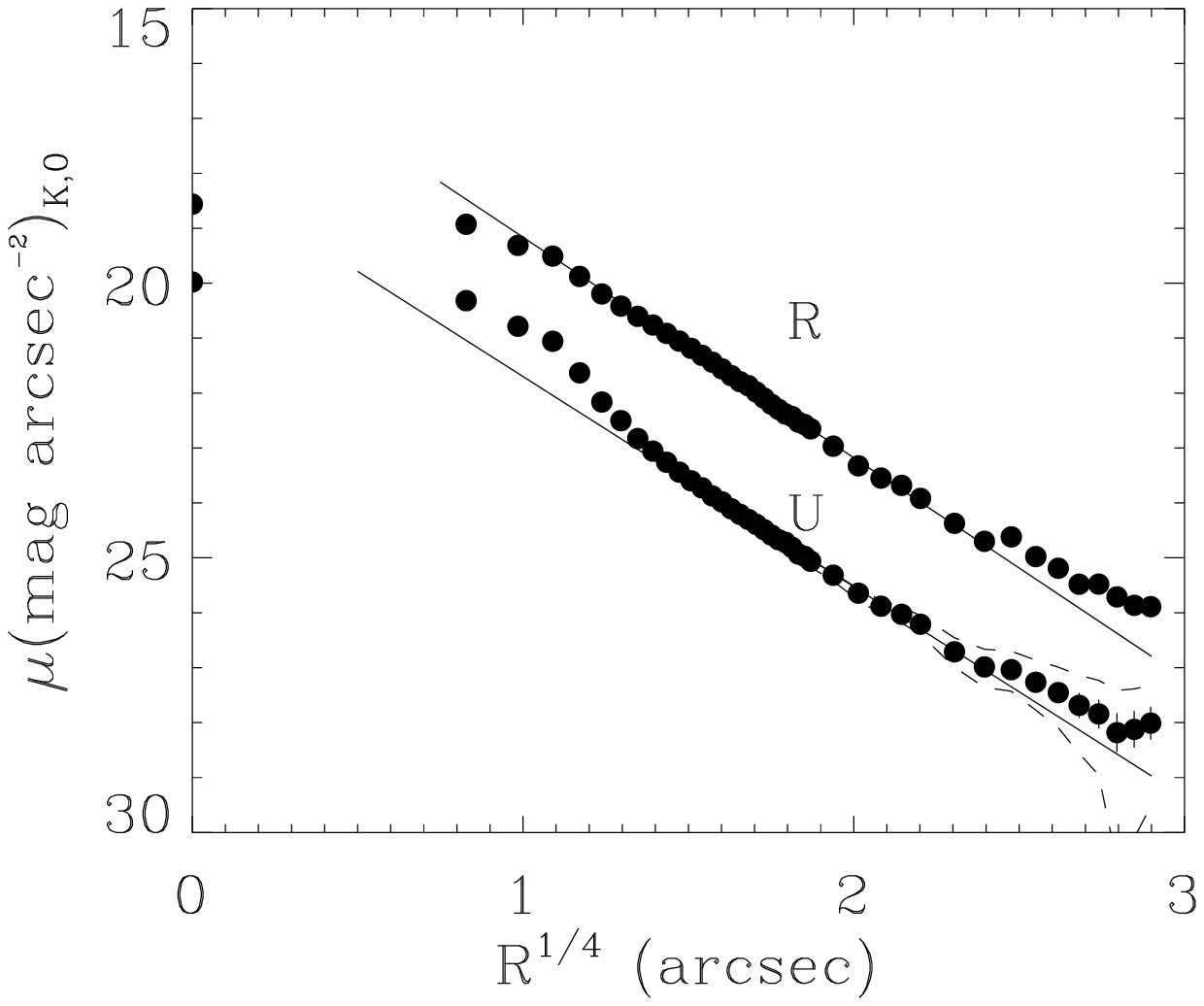}
\end{center}
Figure 1 (left): Correlation between central $U-B$ \und{continuum}
color excess and total cooling rate. Figure 2 (right): $U$ and $R$
 surface brightness profiles for the Abell 1068 CDG.  The solid lines
 represent a $R^{1/4}$-law profile. 
\end{figure}

Unusually blue colors associated with young, massive stars
are often seen in the central $\sim 5-30$ kpc of
cooling flow CDGs (McNamara 1997; Cardiel et al. 1998).  
The likelihood of detecting a blue population correlates strongly
with $\dot m_x$.  This correlation is shown using a $U-B$ color excess
relative to a non-accreting galaxy template in  Figure 1;
it is the strongest evidence linking star formation
to the presence of a cooling flow.  The star formation rates 
associated with the objects in Figure 1 range from $\lae 1-100\msunyr$
(McNamara \& O'Connell 1989, McNamara 1997; Cardiel et al. 1998). 
Beyond the central regions, the spatially
averaged surface brightness profiles usually follow the de Vaucouleurs
$r^{1/4}$ law well into their halos.  If the CDG has the characteristic envelope
of a cD galaxy (Schombert 1987), the  profile rises
above the $r^{1/4}$ law extrapolated outward from the halo.
In Figure 2 I show $U$ and $R$ surface brightness profiles for the CDG
in the distant, $z=0.1386$ cooling flow cluster Abell 1068, 
whose cooling rate is estimated to be 
$\dot m_x\sim 400\msunyr$ (Allen et al. 1995). 
The $U$-band profile rises above the $r^{1/4}$ profile
in the inner several kpc of the galaxy.  Beyond the inner few arcsec,
both the $U$ and $R$ profiles follow the $r^{1/4}$ profile until reaching the
cD envelope at $\mu(R)\simeq 25~{\rm mag.~arcsec}^{-2}$,
where the surface brightness rises above the  $r^{1/4}$ profile
with an amplitude
of $\sim 0.5$ mag.  Apart from the blue core, this surface brightness profile
is typical for cD galaxies in clusters with and without cooling flows
(Porter et al. 1991).  There is little evidence to suggest 
that the {\it average} halo structure and colors of cooling flow galaxies have
recent star formation in excess of what is seen in non cooling flow
galaxies.  The blue inner regions appear to be the result
of accretion concentrated onto the core of a preexisting galaxy,
but evidently not throughout its volume.

\section{Radio Triggered Star Formation}

Most cooling flows harbor luminous $\sim 10^{40-42}~ \ergsec$ emission 
line nebulae extending several to tens of kpc around
the CDG nuclei (Heckman et al. 1989; Baum 1991).  The line
emission and blue optical continuum are usually extended on similar
spatial scales (Cardiel et al. 1998), and the radio and
emission line morphologies and powers are correlated, although
with a large degree of scatter (Baum 1991, but also see Allen 1995). 
The tendency for strong line emission from warm,
$10^4$ K gas to lie along the edges of radio sources is particularly germane to
understanding star formation in these objects.  An early example
was seen in the Abell 1795 CDG (van Breugel et al. 1984), 
and a more striking example 
is seen in H$\alpha$ imagery of the Abell 2597 CDG with the
Hubble Space Telescope (Koekemoer et al. 1999).
Furthermore, the radio jets in Abell 1795
and Abell 2597 bend at roughly 90 degree angles and inflate into radio lobes
at the locations of dust clouds embedded in the
emission-line nebulae (Sarazin et al. 1994; McNamara et al. 1996).  
Their disrupted (i.e. bending) 
radio morphologies are almost surely the result of 
collisions between the radio jets and cold, dense clouds associated
with the line-emitting gas.

At the same time, Abell 2597 and Abell 1795 have bright blue optical continuum
(blue lobes) along their radio lobes (McNamara \& O'Connell 1993;
McNamara 1997), much like the so-called alignment effect seen in
distant radio galaxies (McCarthy 1993). 
That this phenomenon is seen in a relatively small sample of CDGs is
particularly interesting. Unlike distant radio galaxies,
the cooling flow CDGs were selected on the basis of their X-ray
properties, rather than their radio properties. 
Upon their discovery, two models emerged to explain the blue lobes: 
jet-induced star formation
(De Young 1995) and scattered light from an obliquely directed
active nucleus (Sarazin \& Wise 1993; Murphy \& Chernoff 1993; Crawford
\& Fabian 1993).  The scattered light hypothesis predicts the
blue lobe light should be polarized, as is found in many
distant radio galaxies exhibiting the alignment effect (Jannuzi \& Elston
1991; di Serego Alighieri 1989).  $U$-band continuum polarization 
measurements for the Abell 1795
and Abell 2597 CDGs obtained with the KPNO 4m Mayall telescope
gave upper limits of $\lae 6\%$ to the degree of polarization in both
objects, which effectively excluded the scattering hypothesis
(McNamara et al. 1996; 1999).   

Subsequent HST images of both objects resolved the blue lobes
into knots of young star formation (McNamara et al. 1996; Pinkney et al.
1996; Koekemoer et al. 1999).  The HST $R$-band image of Abell 1795's
blue knots are shown against a contour map of the radio source in
Figure 3.  The stellar knots are found along the edges
of the radio lobes and near the collision sites of the radio plasma
and cold gas.  They are not found primarily along the radio jets, 
as would be expected
if the triggering mechanism were shocks traveling transverse to the jet
trajectory, as predicted
in jet-induced star formation models (De Young 1995; Daly 1990, Begelman
\& Cioffi 1989).  The observations suggest that momentum transferred
through direct collisions between the radio
plasma and cold gas clouds may be a more suitable triggering
mechanism.  (D. De Young pointed out that
the strongest shocks would occur at the point of impact, and these
shocks provide a possible triggering mechanism.)

Although star formation at rates of $\sim 10-40 \msunyr$ appears to be 
occurring in these objects, the radio sources may not have triggered
all star formation. In addition to the blue light
along the radio lobes, a more diffuse blue component that
accounts for more than half the blue light is seen.  Therefore,
the radio source may be augmenting star formation in preexisting
star bursts.

\begin{figure}
\begin{center}
\includegraphics*[width=4cm,height=8cm]{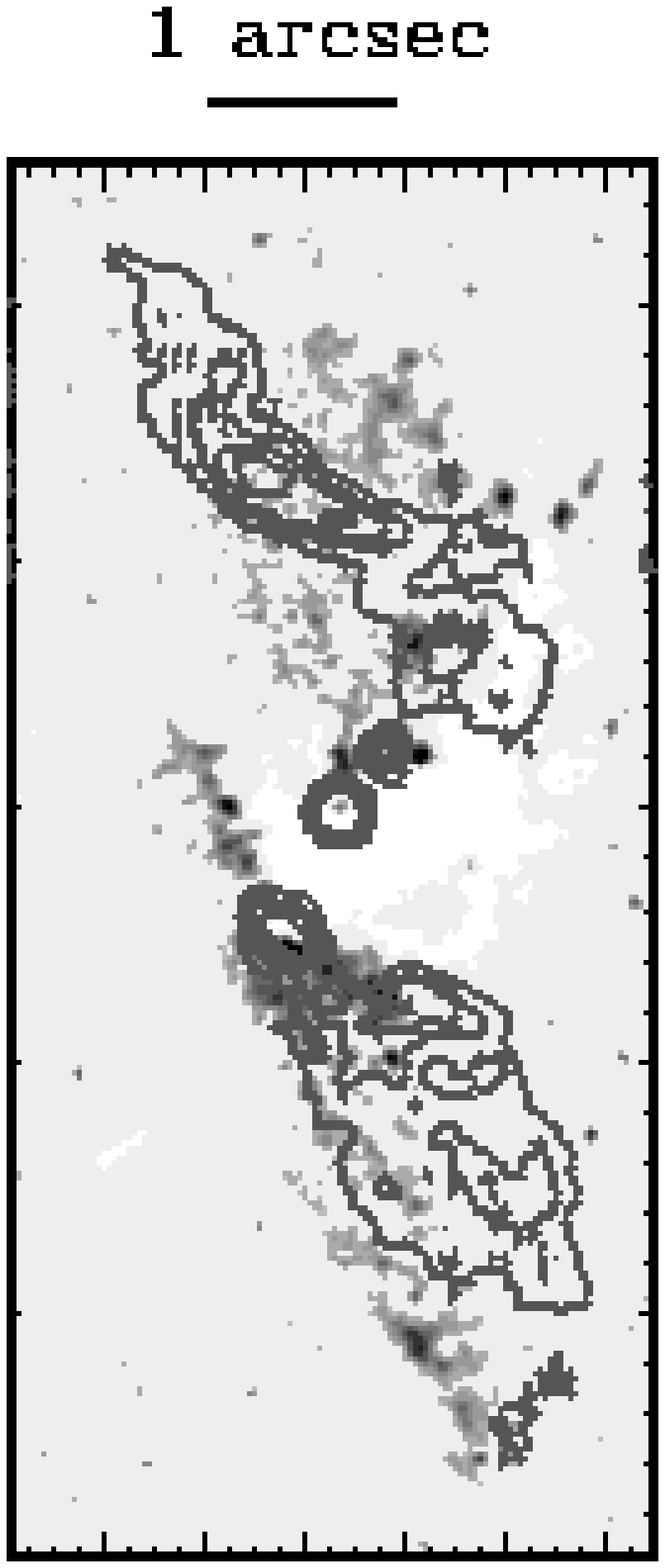}
\includegraphics*[width=7cm,height=7cm]{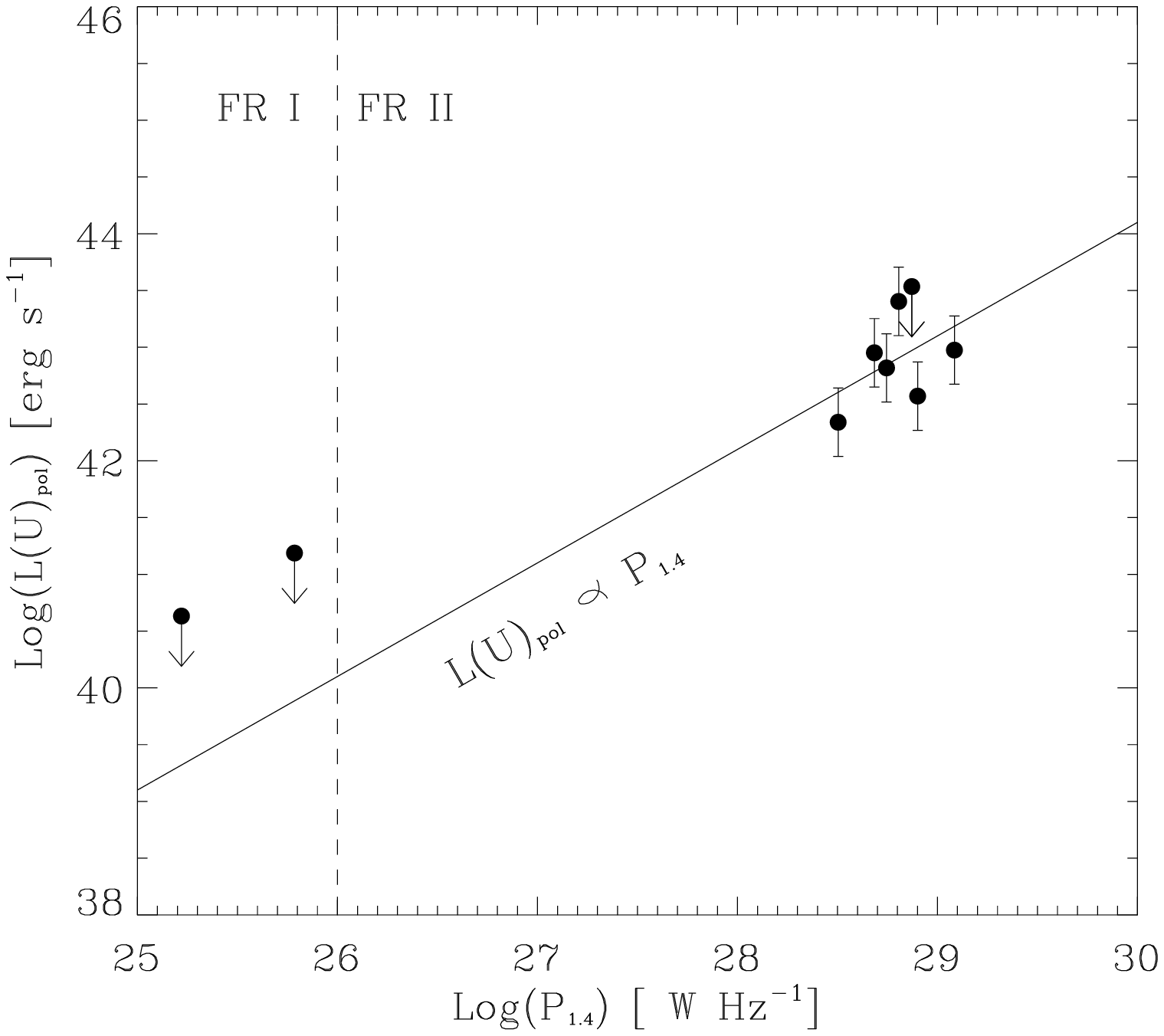}
\end{center}
Figure 3 (left): HST image
in V+R of Abell 1795's blue lobes (greyscale) resolved into knots of
star formation along the 3.6 cm radio lobes (contours).  Figure 4 (right): Radio power plotted against the polarized luminosity for alignment
effect radio galaxies.  The  3C radio
galaxies are grouped to the upper right, and the upper limits to
the polarized luminosity for Abell 1795 and Abell 2597 are to the lower left 
of the plot.  The solid line, normalized to the median of the 3C points,
represents $L(U)_{\rm pol}\propto P_{\rm rad}$.  The vertical dashed 
line indicates approximately the transition between FR I and FR II
radio luminosities. 
\end{figure}
\section{A Burst Mode of Star Formation in Cooling Flows}
\label{Properties}

Tracing the history of a stellar population, even in isolation, is difficult.
The problem is further complicated when the population is embedded
in a bright background galaxy. The blue lobes in
the Abell 1795 and Abell 2597 CDGs are the first clear-cut evidence for 
a burst mode of star formation in cooling flows.  The blue
lobes cannot be old because the
the alignment between the radio and optical structures can 
last only a fraction of
the radio source lifetime and the stellar diffusion time scale,
both $\sim 10^7$ yr.  Additional evidence supporting a burst
mode of star formation in cooling flows has accumulated in
recent years. Cardiel et al. (1998) have argued using the Mg II 
absorption line index,
the 4000 \AA\ break, and far UV colors that short duration
bursts ($\lae 10^7$ yr) or constant star formation with
ages $\ll 1$ Gyr best fit Bruzual model isochrones.
While acknowledging the large uncertainties in the population isochrones,
a burst mode of star formation is unexpected 
in simple, continuous cooling flow models (e.g. Fabian 1991).
If star formation is indeed being fueled by cooling flows, 
it would seem that gas is not accreting continuously.
Transient sources of fuel, such as mergers or stripping, 
may also be contributing.

\section{Are CDGs in Cooling Flows Low Radio Power Siblings of High Redshft Radio Galaxies?}

The premise that blue lobes are sites of star formation is
supported by several facts.  The absence of a polarized signal from the
blue lobes effectively excludes the scattered light hypothesis.
Synchrotron radiation can be excluded by the absence of a 
detailed correlation between the radio source and blue lobes,
and the nebular continuum is insufficiently strong to account for the
blue color excesses.  However, Balmer absorption is seen in
the spectra of some objects (Allen 1995), and the emission line
luminosities and H II region characteristics
are often consistent with powering by young stars (Shields \&
Filippenko 1990; Voit \& Donahue 1997), so star formation is 
almost certainly the primary source of the color excesses in CDGs.
The situation is more complex in the high redshift 
powerful radio galaxies (HzRGs) exhibiting the alignment effect.
The aligned optical continuum in HzRGs 
is often strongly polarized, which has been interpreted as
the signature of scattered light from an obliquely-directed active nucleus 
(di Serego Alighieri et al. 1989; Jannuzi \& Elston 1991).  
In Figure 4, I plot our polarized flux upper limits 
for the blue lobes in Abell 1795, Abell
2597, and the alignment regions of several
HzRGs against rest frame 20 cm radio power
(see McNamara et al. 1999).  The polarized fluxes are measured in
the rest frame $U$-band, and can be compared directly.  Although the
HzRGs are 2--3 orders of magnitude more powerful in their radio and
polarized fluxes, a linear extrapolation
downward between radio power and polarized flux from the mean
HzRG value to the cooling flows would predict a lower
polarized flux than is observed.
Assuming similar host galaxy properties and scattering
environments in both types of object, 
and further assuming the polarized flux scales approximately in
proportion to radio power (see McNamara et al. 1999), 
at the precision of our measurements,
we should not have detected a polarized flux in Abell 2597 and Abell 1795.
In addition, it would seem that the polarized fluxes
of  HzRGs generally account for a large but incomplete fraction of 
the blue light, and occasionally unpolarized star light dominates
(e.g. van Breugel et al. 1998).
It is possible then that the blue lobes in cooling flows
and the alignment effect in powerful radio galaxies are similar
phenomena.  But while starlight dominates the aligned continuum in lower
radio power CDGs, scattered light dominates in HzRGs
owing to their more powerful nuclei (McNamara et al. 1999).

\section{An Analysis of New Imagery for the Abell 1068 CDG}
\label{A1068}

In this section I discuss new optical imagery of the Abell 1068
central cluster galaxy.  The data provide new clues
to the relationship between star formation and the radio
source, and raise new questions regarding the mechanism fueling star formation. 
$U$-band CCD imaging is the most sensitive means of isolating
and studying the bluest galaxy populations from the ground.
The blue populations in CDGs often contribute
more than half of the central $U$-band
light, while the fraction decreases to $\sim 10\%$ or less
in the $R$ and $I$ bands. The blue
populations can therefore be isolated by modeling and
subtracting the background galaxy leaving the blue regions
in residual.  By doing so in two or more pass bands,
intrinsic colors of the blue population can be estimated.

\begin{figure}
\begin{center}
\includegraphics*[width=6cm]{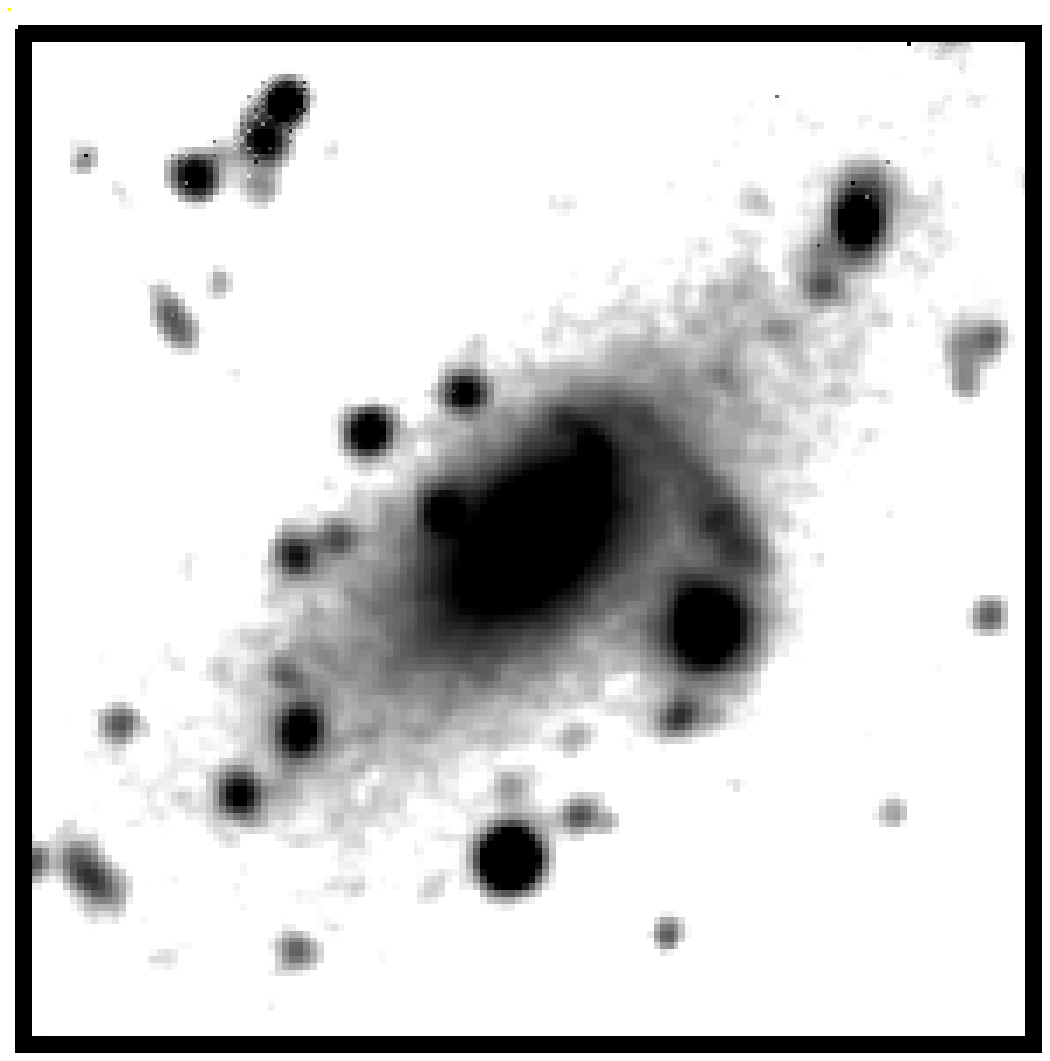}
\includegraphics*[width=6cm]{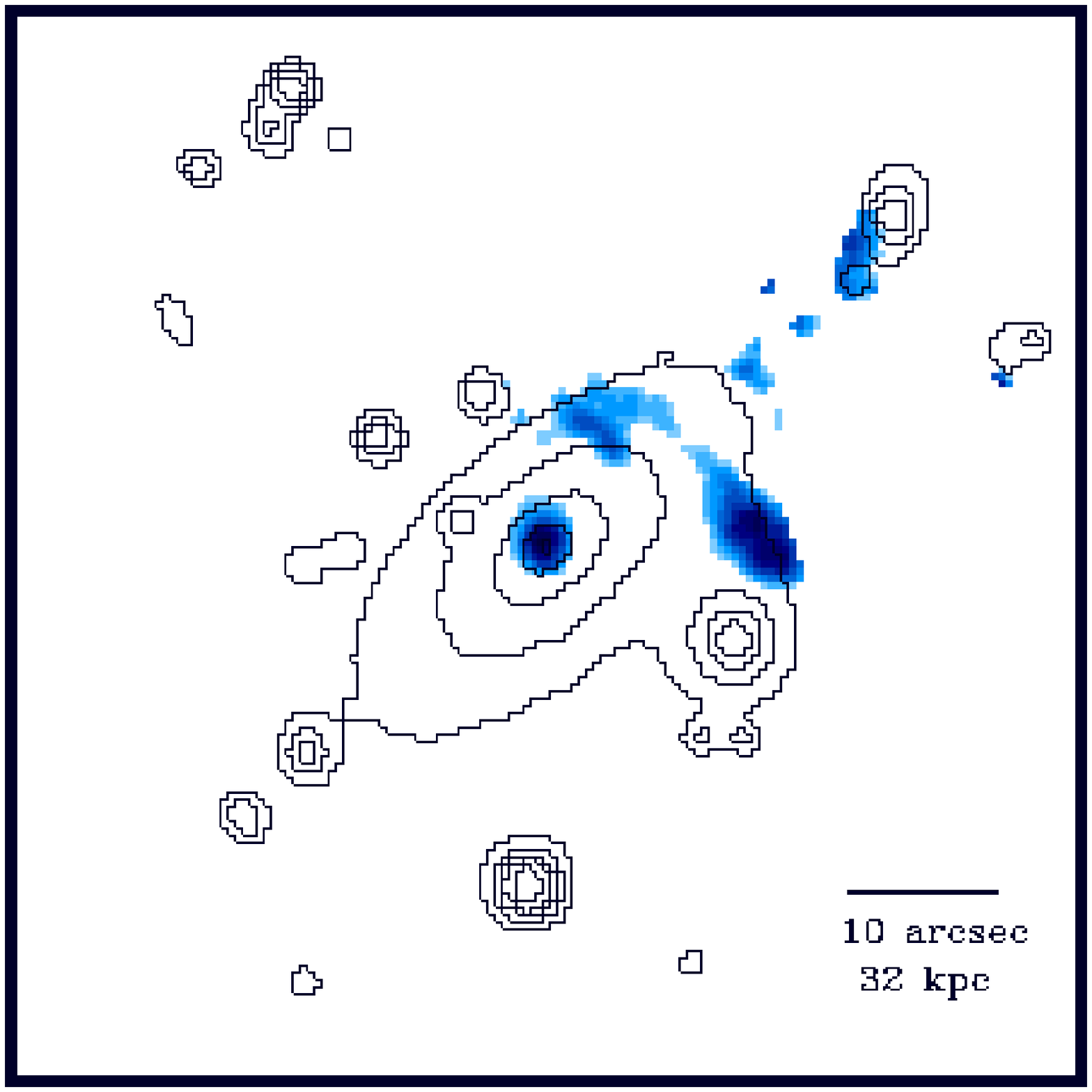}
\includegraphics*[width=6cm]{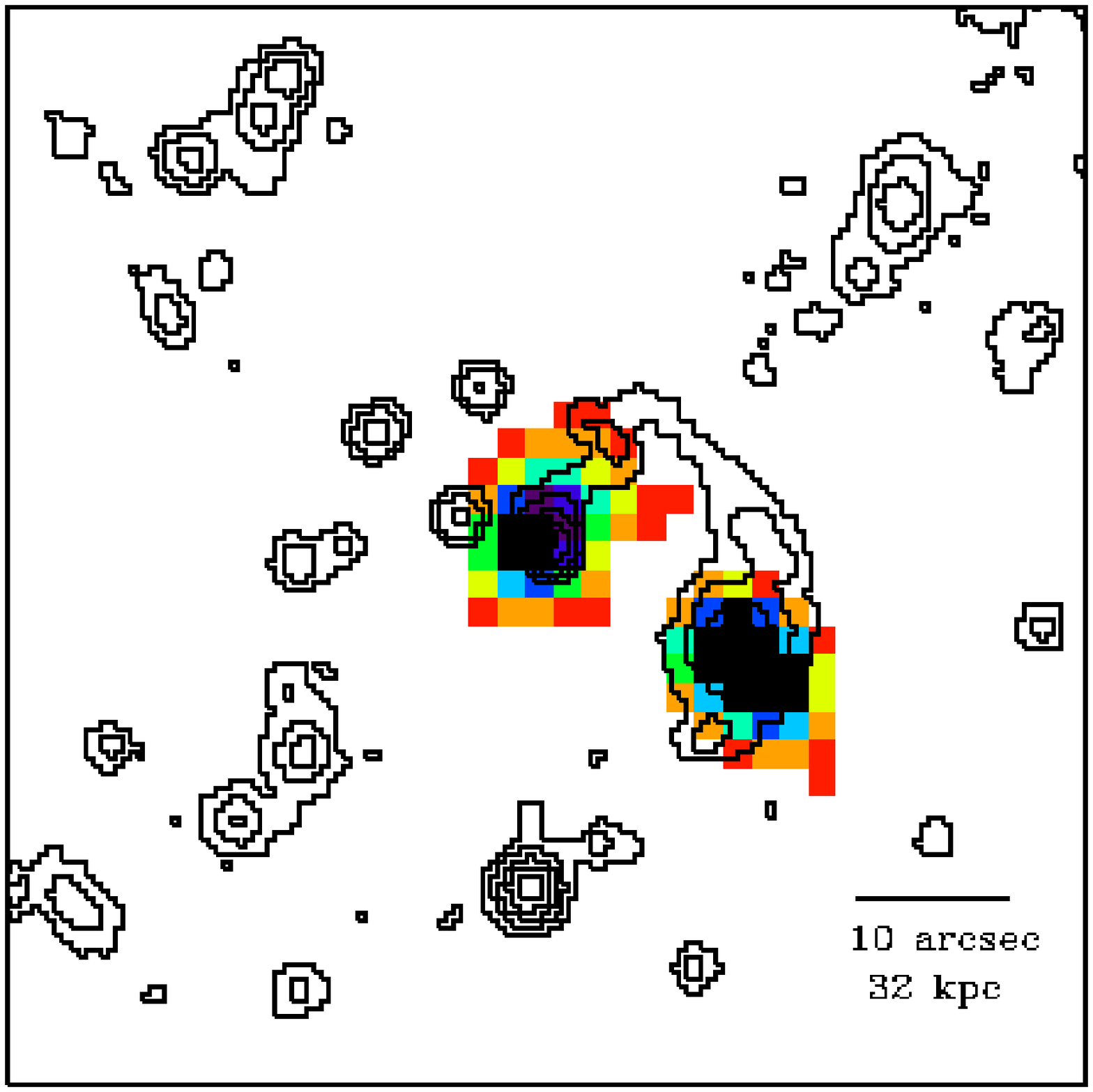}
\includegraphics*[width=6cm]{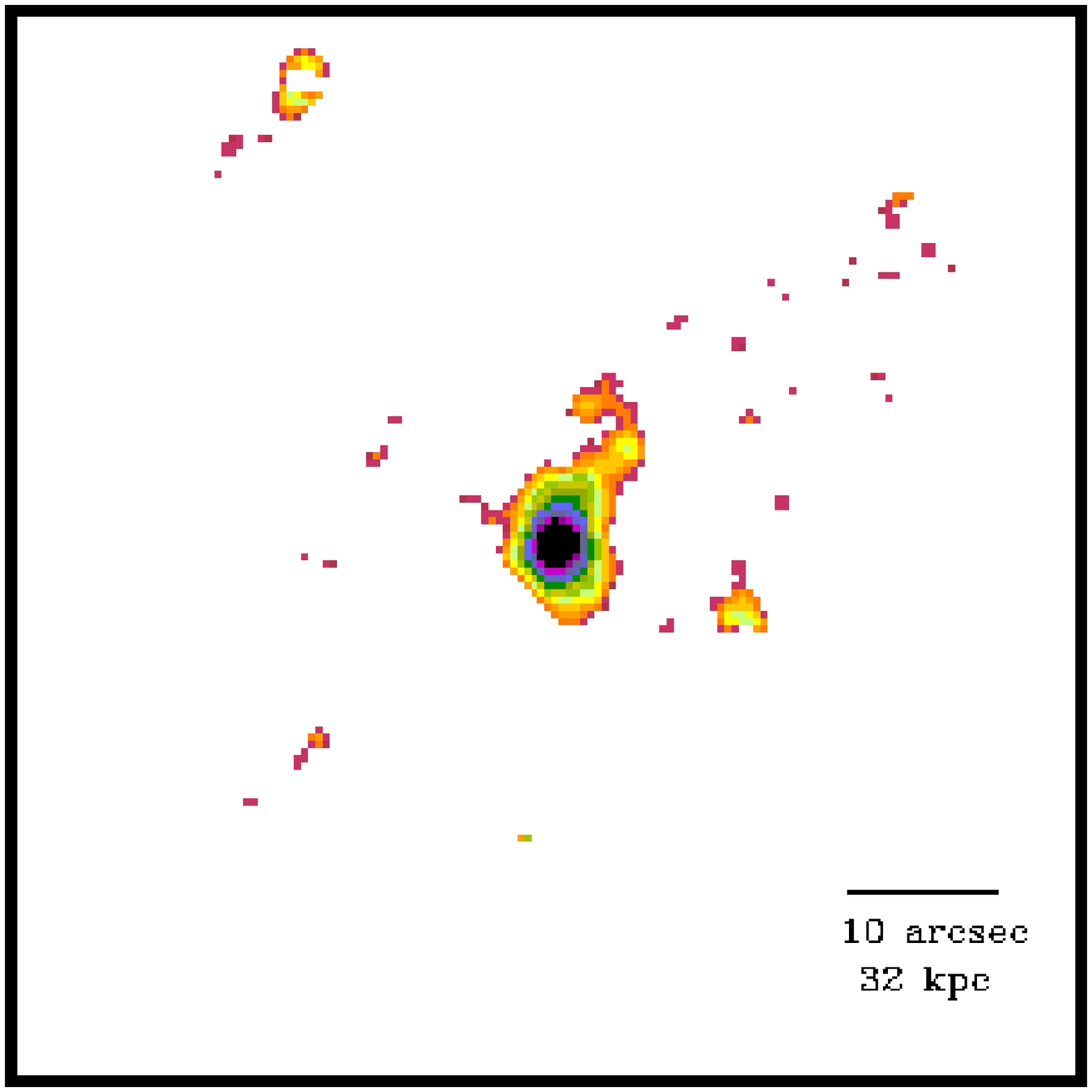}
\end{center}
Figure 5: Imagery of Abell 1068: $U$-band image (upper left); $U-R$ color
map (grayscale) superposed on $R$-band contours (upper
right); $U$-band contours, after subtracting a smooth $U$-band
model CDG galaxy, on the 20 cm FIRST radio
grayscale image (lower left); H$\alpha$ map (lower right).  
The panels are registered to the same scale; 
north is at top and east is to the left. 
\end{figure}

I applied this procedure to the $z=0.1386$ Abell 1068
CDG, one of the most distant and largest
cooling flows ($\dot m_x\sim 400\msunyr$) discovered in the 
$ROSAT$ All Sky Survey (Allen et al. 1995).  It is also
one of the bluest CDGs in my sample.  Figure 5 presents 4-panels
showing the $U$-band image to the upper left, a $U-R$ color
map (grayscale) superposed on $R$-band contours to the upper
right, $U$-band contours, after subtracting a smooth $U$-band
model CDG galaxy, on the 20 cm FIRST radio
grayscale image ($FWHM=5.4$ arcsec), lower left, and an H$\alpha$ map,
lower right.  The panels are registered to the same scale;
north is at top and east is to the left.  Gray regions in the
color map are abnormally blue.

Several features are noteworthy.  First, the central region
within a 13 kpc diameter is $\sim 0.5-0.9$ mag bluer than normal.
The nuclear colors, after K correction, range between
$(U-R)_{\rm K,0}\simeq 1.5-2.3$ (the foreground reddening is negligible).
An arc of blue light lies 8 arcsec (25 kpc) in projection to the north-west 
of the nucleus, and a large wisp or
arc of blue light extends  to the south-west, until meeting a
bright blue patch of light 13 arcsec to the east of the nucleus,
and about 8 arcsec to the north of the bright neighboring galaxy
to the south-west of the nucleus. This feature is nearly as
blue as the nucleus with $(U-R)_{\rm K,0}\simeq 1.6$. Finally, several
blue knots, 15--30 arcsec north-west of the nucleus, appear 
along a line between the nucleus and 
a disturbed galaxy 35 arcsec to the north-west
of the nucleus.  The remaining colors of the off-nuclear features range from 
$(U-R)_{\rm K,0}\simeq 2.0$ to the normal background color
$(U-R)_{\rm K,0}\simeq 2.4$.

After subtracting a model galaxy from the $U$ and $R$ CDG images, 
I find an intrinsic nuclear blue population color 
$(U-R)_{\rm K,0}\sim -0.2$. 
This color is consistent with Bruzual-Charlot
population model colors for a $\sim 10^7$ yr old burst population
or continuous star formation for $\sim 0.1$ Gyr.  The colors
are bluer than expected for star formation in a cooling flow 
that has been accreting continuously for $\gae 1$ Gyr.  
The accretion population's luminosity mass is
$\sim 2\times 10^8\msun$, which
would correspond to a star formation rate of $\sim 80 ~\msunyr$.
The off-nuclear colors, being a few
tenths of a magnitude redder than the nuclear colors, are consistent
with a several $10^7$ yr old burst or continuous star formation
for $\lae 1$ Gyr.  

The off nuclear blue regions are apparently not in dynamical
equilibrium.  They appear to be stripped debris, possibly
from the bright neighboring galaxies to the north-west
and south-west of the nucleus.  The disturbed appearance of
the north-west galaxy's $R$-band isophotes support the
stripping hypothesis.  The blue regions are
considerably bluer than their putative parent galaxies, 
which would be consistent with blue material
being composed primarily of young stars that formed out of cold material
stripped from the galaxies.

\subsection{Radio Triggered Star Formation in Abell 1068?}
\label{Fuel}

Both the Abell 1068 CDG and the bright galaxy
to the south west of the CDG are radio sources. Each have radio
powers of $\sim 8.5\times 10^{24}$ W/Hz, which are
typical for FR I radio sources.  In addition, the nucleus
is embedded in a luminous emission line nebula with an H$\alpha$ 
luminosity $\gae 2\times 10^{42}~\ergsec$ (Allen et al. 1992).
Although only a low resolution radio map is available,
the radio source appears extended to the north-west in the same
direction as a tongue of H$\alpha$ emission extending from the nucleus.  
Both the radio source and the tongue of H$\alpha$ emission terminate
8 arcsec (25 kpc) to the north-west of the nucleus at the location
of the bright blue arc.  Such a close spatial relationship between
the radio source, nebular emission, and knots of star formation
are common in powerful radio galaxies in general, and in cooling
flows in particular.  It is tempting to speculate that, with high
resolution radio maps in hand, the radio and optical
morphologies will again be consistent with radio triggered star
formation in the blue arc to the north-west, much like
Minkowski's Object (van Breugel et al. 1985).

\section{The Fueling Mechanism}
\label{Fuel}

The origin of the material fueling star formation is
of fundamental interest.  A cooling flow origin is supported by the
correlation between central blue color excess in
CDGs and the
cooling rate of the intracluster gas, derived independently
from X-ray observations, shown in Figure 1
(e.g. McNamara 1997; Cardiel et al. 1998).
Were major galaxy mergers supplying the fuel, 
this correlation would be difficult to
to explain.  I would then expect CDGs experiencing significant bursts
of star formation to be observed
with equal frequency in cooling flow and non-cooling flow
clusters alike, but they are not.  Nonetheless, the evidence supporting
periodic bursts of star formation implies an intermittent
source of fuel.
Ram pressure stripping of cold gas from neighboring cluster
galaxies may be such a source of fuel, and might account for the
$\dot m_x$--blue color correlation.  The cooling rate 
$\dot m_x\propto\rho_{\rm gas}^2$, and the ram pressure force on a
parcel of gas is $\rho_{\rm gas}v^2$.  Therefore,
the dense cooling flow regions provide a large stripping cross 
section capable of 
sweeping cold, dense molecular gas from cluster
dwarf galaxies and spirals, which would rain onto the parent
CDG.  Abell 1068 may be a case in point,
as might the Abell 1795 CDG (McNamara et al. 1996).

\section{Cooling Flows and the Chandra X-ray Observatory}

As I wrote this article, Chandra was launched and began sending
astonishingly crisp images of cosmic X-ray sources.  During the
next few years, many of Chandra's targets will be clusters of
galaxies, and the cooling flows promise some of
the most interesting and productive cluster science.  
Their bright cores--the characteristic signature of a cooling flow--afford 
Chandra the opportunity to take full advantage of its nearly
perfect, half arcsecond mirrors.  For the first time, we will 
be capable of mapping structure in the X-ray-emitting gas
on angular scales smaller than the radio sources and star 
formation regions.  The temperature and density maps on these small scales
will provide local cooling rates that can be compared directly
to optically-derived star formation rates.  Perhaps more than any other X-ray
telescope planned or in queue, Chandra will advance
our understanding of the dynamical and thermal state of cluster
cores, which hopefully will bring the long-standing cooling flow
problem to resolution.

\section{Summary}

$\bullet$ Unusually blue colors associated with young, massive stars
frequent the central regions of cooling flow CDGs.  The probability of
detecting a blue population increases sharply with $\dot m_x$ derived
from X-ray observations.

$\bullet$ Star formation in cooling flows apparently occurrs
in repeated, short duration ($\lae 1$ Gyr) bursts, not
continuously as would be expected in standard cooling flow models.

$\bullet$ Bursts of star formation are often triggered by the radio sources.

$\bullet$ Cold material stripped from neighboring galaxies may feed the
the radio source and fuel some star formation in CDGs.

\end{document}